\documentclass[lettersize,journal]{IEEEtran}
\usepackage{amsmath,amsfonts}
\usepackage{algorithmic}
\usepackage{algorithm}
\usepackage{array}
\usepackage[caption=false,font=normalsize,labelfont=sf,textfont=sf]{subfig}
\usepackage{textcomp}
\usepackage{stfloats}
\usepackage{url}
\usepackage{verbatim}
\usepackage{graphicx}
\usepackage{cite}
\usepackage{tikz}
\usepackage{multirow}
\usepackage{tabularx}
\usepackage{mathrsfs}
\usepackage{bm}
\usetikzlibrary{shapes,arrows,positioning}
\begin{document}

\title{On Architectures for Combining Reinforcement Learning and Model Predictive Control with Runtime Improvements}

\author{Xiaolong Jia*, Nikhil Bajaj*\\ *Department of Mechanical Engineering, Stanford University \\**Department of Mechanical Engineering and Materials Science, University of Pittsburgh}


\maketitle

\begin{abstract}
Model Predictive Control (MPC) is a robust control algorithm extensively utilized in domains such as autonomous driving and robotics. MPC addresses control as an optimization problem at each time step, and can manage dynamic and input constraints by incorporating them into the constraints of the optimization. Considerable computational demand is inherent to many of the variations of MPC, as the optimization must be executed repeatedly at each control interval. In practical control applications, discrepancies between the theoretical model and the actual dynamics of the system often lead to suboptimal performance. Unlike the mostly offline design approach of MPC, Reinforcement Learning (RL) develops an optimal control policy through iterative online data acquisition. RL typically optimizes its control policy using environmental rewards and observations without comprehensive knowledge of the system dynamics \cite{Lin2023ReinforcementSystems}. This study integrates a model-free reinforcement learning-based strategy to alleviate challenges brought about by parameter variations and to reduce the computational burdens of MPC by employing Neural Networks (NN) to approximate the MPC algorithm. By merging the online, data-driven adaptation characteristics of RL with the offline optimization capabilities of Neural Network approximated Model Predictive Control (NNMPC), the proposed reinforcement learning-assisted model predictive control (RLMPC) effectively mitigates some of the runtime limitations associated with MPC.
\end{abstract}

\begin{IEEEkeywords}
Model predictive control (MPC), Neural Network (NN), reinforcement learning (RL), parameter fluctuations, policy gradient.
\end{IEEEkeywords}

\section{Introduction}
Model predictive control (MPC) is a powerful control algorithm widely used in areas such as automatic driving and robotics. Using a predictive model to optimize over different combinations of inputs over time, MPC solves a finite horizon optimal control problem to achieve good trajectory tracking while handling constraints. The performance of the predictive model, which is the core of model predictive control, depends significantly on the accuracy of the parameters of the real system. Most research assumes close alignment between predictive models and actual systems. However, real-world scenarios reveal discrepancies due to factors like manufacturing inconsistencies, environmental changes, and time-related wear\cite{Martin2017SensitivityDrives}. These variations challenge control laws derived from fixed models, potentially degrading performance, stability, and robustness. To address this, approaches such as adaptive control\cite{Bujarbaruah2018AdaptiveKeeping}, robust control\cite{Borrelli2017PredictiveSystems}, and online parameter learning control\cite{Gros2022LearningGuarantees} are employed in MPC.

Machine learning has been considered for its potential in control areas for decades, advancing our ability to describe complex systems from observational data, in complement to first principles modeling \cite{Brunton2019Data-DrivenControl}. All of machine learning revolves around optimization, including regression and model selection frameworks that aim to provide parsimonious and interpretable models for data\cite{Hardle2019AppliedAnalysis, Brunton2019Data-DrivenControl, Wong2018RecurrentManufacturing}. The integration of neural networks (NN) into MPC by fitting the outcome of the optimization solutions across the state space and using those solutions to apply input to the plant maintains the performance of MPC, but also reduces the large computational cost of receding horizon control to a considerable extent \cite{Karg2020EfficientLearning}. While it is much less computationally intense to implement, this Neural Network-based Model Predictive Control (NNMPC) is only an approximation of MPC, and so it retains MPC's disadvantages. It can be considered a variation of explicit model predictive control (eMPC). MPC/eMPC can not only take as input the current state and reference, but also previewed signals (e.g. future reference values) within the prediction horizon. While can improve control performance, when approximating the MPC by a NN, a larger input layer for the neural network is required. The number of neurons in the input layer is based on the dimensionality of the input data (state and reference space), directly impacting the number of weights and computations required in the first hidden layer. A larger input layer increases the computational load. In addition, this greatly increases the data set required to span the reference and state space to approximate the MPC solutions.

Reinforcement learning (RL) is an online learning process that refines its control policy (some control law implemented with logic and/or parameters) over training time, aiming at achieving improved control performance measure. In contrast with the explicit NNMPC’s offline optimization approach, Reinforcement Learning (RL) generates an optimal control policy through repetitive online data observation. Without requiring the actual system dynamics, RL can its control policy using reward and observations from the environment \cite{Lin2023ReinforcementSystems}. Recent research efforts have utilized RL to improving MPC. An illustrative approach combines model predictive control (MPC) with reinforcement learning (RL) via policy iteration (PI), where MPC serves as a policy generator, and the RL method is employed to assess the policy -- this falls into the category of policy-iteration approaches. The resultant value function is then used as the terminal cost in MPC, thereby enhancing the produced policy. This method benefits from eliminating the necessity for the conventional MPC's offline design framework of terminal cost, auxiliary controller, and terminal constraint cite{Lillicrap2015ContinuousLearning,Lin2023ReinforcementSystems}.

This paper combines the approximation capabilities of neural networks with the model-free reinforcement learning approach, integrating an actor-critic method into an enhanced NNMPC framework. This allows for a reduction in the input layer size and computational cost without compromising NNMPC's control performance. The model-free strategy, grounded in reinforcement learning, aims to address the challenges arising from parameter variations encountered when employing model-based NNMPC trained through offline supervised learning.

Our strategy presents a straightforward way to create predictive reinforcement learning (RL) policies, differing from MPC-driven policy search methods. By leveraging the Deep Deterministic Policy Gradient (DDPG) algorithm, we learn and approximate an MPC for use within the actor network with an NNMPC approach. Prediction steps from reference trajectories are processed and used as observational data, helping the actor network understand NNMPC's core characteristics. This enables the refinement of the entire control approach via the policy gradient. We allow MPC to handle most of the control task, while the RL strategy assumes a secondary role. This distribution maintains training safety within certain uncertainty boundaries because the RL contribution is conservatively bounded. We demonstrate this on a rotary inverted pendulum system to effectively address parametric uncertainties with the RL and NNMPC.

\section{Preliminaries}
This section provides an overview of the deficiencies in Model Predictive Control (MPC) arising due to parameter variations, alongside pertinent background information relevant to our methodology. 

\subsection{Model Predictive Control and Suboptimal Performance}
The classical MPC assumes a linearized dynamic model and solves a quadratic programming problem.  
Consider a linear time-invariant discrete-time system given by:

\begin{equation}
      \bm{x}(k+1) = A\bm{x}(k) + B\bm{u}(k)
\end{equation}

The objective of MPC is to minimize the following cost function over a finite prediction horizon $N$, and the optimization problem can be formulated as \cite{Wang2009ModelMATLAB}:
\begin{align}
    \min_{U} \ & J(U)=\sum_{k=0}^{N-1} \left( \|\bm{x}(k) - \bm{r}(k)\|^2_Q + \|\bm{u}(k)\|^2_R \right) \label{Goal}\\
    \text{s.t.} \ & \bm{x}(k+1) = A\bm{x}(k) + B\bm{u}(k), \ k = 0,1,...,N-1 \\
    & \bm{x}_{\text{min}} \leq x(k) \leq \bm{x}_{\text{max}}, \ k = 1,2,...,N \\
    & \bm{u}_{\text{min}} \leq u(k) \leq \bm{u}_{\text{max}}, \ k = 0,1,...,N-1
\end{align}

The optimization is subject to system dynamics and constraints on states and control inputs. The Q and R are weight matrices, similar to other optimal control approaches like linear quadratic regulators (LQR). $Q=diag([q_{1},q_{2},...,q_{n}])$, and for each input: $R=diag([r_{1},r_{2},...,r_{m}])$. $U = \{u(0), u(1), ..., u(N-1)\}$ is the sequence of control inputs. For each time step, the MPC only applies $u(0)$ to the plant, and then waits for the next time step before performing optimization again, moving the prediction sequence one time step ahead. This is the principle of receding horizon control. Thus, the MPC control algorithm computes, at each time step $k$:

\begin{equation}
      \bm{u}_{k}=f_{MPC}(\bm{x}_{k},W_{k})
\end{equation}

where $x_{k}$ is the current state vector at time step $k$, and $W_{k}$ is a matrix of reference steps for all states within the prediction horizon. In the cases considered here, the MPC assumes linearized but exact discrete-time dynamics to generate the inputs. However, in real world dynamics, model parameter and structural errors exist - and so the theoretical model may not agree with the actual dynamics. This leads to to suboptimal performance due to the prediction errors.

Optimal control theory deals with finding a control law for a given system such that a certain optimality criterion is achieved. Optimization problems usually include minimizing the cost, and the objective is defined in \eqref{Goal}. Considering the uncertain system:

\begin{equation}
      \bm{x}(k+1) = A({w}^{p}(t))\bm{x}(k) + B({w}^{p}(t))\bm{u}(k)
\end{equation}
where ${w}^{p}\in\mathbb{R}^{{n}_{p}}$ represents the parameter variations.

We can define a ``ground truth'' cost function:
\begin{align}
    \mathcal{{J}_{W}}\ &=\sum_{k=0}^{N-1} \left( \|\bm{x}(k) - \bm{r}(k)\|^2_Q + \|\bm{u}(k)\|^2_R \right) \label{ActualCost}\\
    \text{s.t.} \ & \bm{x}(k+1) = A({w}^{p}(t))\bm{x}(k) + B({w}^{p}(t))\bm{u}(k)\\
    & \bm{u}(k)=\text{arg min } J(U)  \\
    & k = 0,1,...,N-1
\end{align}

The MPC's offline design prevents accounting for parameter variation in the actual plant. The control difference between the ideal model and real world model can be written as :

\begin{equation}
      \bm{x}_{e}= \hat{f}_{{w}^{p}}(\bm{x}(k),\bm{u}(k))-{f}_{ideal}(\bm{x}(k),\bm{u}(k))
\end{equation}

The resulting difference in output performance $\bm{x}_{e}=\bm{\hat{x}}-\bm{x}$ indicates that since the actual model is unknown, the offline control law may no longer be able to handle the constraint $\bm{x}_{\text{min}} \leq x(k) \leq \bm{x}_{\text{max}}$, leading to suboptimal performance. Certainly, stability is not guaranteed, subject to the robustness of the particular MPC design. In addition to stability issues in the controlled system, the overall control strategy may become less robust to disturbances and noise. In addition, when one considers the ``ground truth'' cost function \eqref{ActualCost}, the cost is likely not minimized.

\subsection{Neural Network Approximated MPC (NNMPC) and Reinforcement Learning}
The objective of neural network function approximation is to find a mapping function \( f \) approximated by the neural network that mimics a target function \( g \) as closely as possible. Given training data pairs \((x, y)\), where \( y \approx g(x) \), the neural network adjusts its internal parameters to learn the function \( f(x) \) such that \( f(x) \approx y \). A neural network (specifically, a multilayer perceptron (MLP) consists of:
\begin{itemize}
    \item Input Layer: Receives input \( \mathbf{x} \in\mathbb{R}^n \).
    \item Hidden Layers: Each layer \( l \) transforms its input \( \mathbf{h}_{l-1} \) into an output \( \mathbf{h}_l \) using weights \( \mathbf{W}_l \) and biases \( \mathbf{b}_l \) followed by an activation function \( \sigma \), i.e., \( \mathbf{h}_l = \sigma(\mathbf{W}_l \mathbf{h}_{l-1} + \mathbf{b}_l) \).
    \item Output Layer: Produces the final approximation \( \hat{y} = \mathbf{W}_{out} \mathbf{h}_L + \mathbf{b}_{out} \), where \( L \) is the last hidden layer. 
\end{itemize} To train the network, we go through several steps:
\begin{itemize}
    \item Forward Propagation: Input data passes through the network to produce an output.
    \item Loss Calculation: The network's output is compared to the true values to compute error using a loss function.
    \item Backpropagation: The error is used to calculate gradients for each network weight.
    \item Weight Update: Weights are adjusted using optimization algorithms like gradient descent to minimize the loss. 
\end{itemize} 

We refer here to neural network approximated model predictive control (NNMPC) as an approach that approximates the MPC solution across the state an input space, allowing this surrogate to be used in place of online optimization. This can be considered to belong to a larger family of explicit model predictive control algorithms. Ultimately, they trade off online computation cost against memory footprint. The forward evaluation of a simple MLP is typically far less computationally expensive than online optimization. Another potential advantage of using NN to approximate the MPC algorithm (rather than some other approximation method) is its potential for generalization based on the training data set. Achieving good performance on new, unseen data, can be confirmed through validation techniques and ensure the model does not just memorize training data but learns to generalize from it. This does require some careful tuning of hyperparameters (e.g. learning rate, network size). 

The structure of the NN is crucial for successful training of an NNMPC. The architecture should be complex enough to capture the necessary patterns but not so complex that it becomes prone to overfitting or leads to large computational cost. The trained NN should be able to approximate the control performance of MPC while reducing the time needed to perform a computation of the needed control effort. Having a large and representative dataset that accurately captures the underlying patterns and variations of the target function or phenomenon is essential. Clean, well-preprocessed data free from biases and errors also enhance the network's ability to learn effectively.

The training of NNMPC is a supervised learning process, and as discussed before, MPC is an offline algorithm. This implies that for NNMPC as well, the previously discussed potential control performance issues due to modeling problems exist here in implementation as well.

Here, we propose that one means to address some deficiency caused by real world model parameter variation could be RL. Deep Deterministic Policy Gradients (DDPG) is an algorithm in reinforcement learning that combines the benefits of policy gradient methods and deep Q-learning. Generally, the algorithm initializes the critic network \(Q(s,a|\theta^{Q})\) and actor network \(\mu(s|\theta^{\mu})\) with weight \(\theta^{Q}\) and \(\theta^{\mu}\). Then a copy of the critic network \(Q'(s,a|\theta^{Q'})\) and actor network \(\mu'(s|\theta^{\mu'})\) is created to calculate the target values. For each training episode, the DDPG algorithm initializes a random noise $\mathcal{N}$ and select action at time step t \(a_{t}=\mu(s_{t}|\theta^{\mu})+\mathcal{N}\) according to current policy and the exploration noise. Through experience replay and policy gradient, the target networks are updated using reward from the environment to get an optimal policy \(\mu^{*}(s|\theta^{\mu^{*}})\) \cite{Lillicrap2015ContinuousLearning}:

\begin{equation}
    \bigtriangledown_{\theta^{\mu}}J\approx\frac{1}{N}\sum_{i}\bigtriangledown_{a}Q(s,a|\theta^{Q})|_{s=s_{i},a=\mu(s_{i})}\bigtriangledown_{\theta^{\mu}}\mu(s|\theta^{\mu})|_{s=s_{i}}
\end{equation}
where $i$ represents a random minibatch of M transitions.

In the context of a control system, DDPG helps in learning a control policy that maps the states of the system model and the references in target trajectory to the continuous actions like system inputs that guide the system to achieve desire targets by exploring and adapting to the environment, and optimizing its performance based on received rewards.

\section{Method}
\subsection{Dynamic Model} In this experiment, a rotary inverted pendulum (model QUBE-Servo 2 by the Quanser), was used in implementing the RLMPC algorithms. The model dynamics are  nonlinear with the state variables \(\bm{x} = [\theta,\alpha,\dot{\theta},\dot{\alpha}]'\), where \(\theta\) and \(\dot{\theta}\) is the angular position and the angular velocity of the rotary arm, \(\alpha\) and \(\dot{\alpha}\) refers to the angular position and the angular velocity of the inverted pendulum. The initial states \(\bm{x} = [0,0,0,0]'\) suggests a physical state that the pendulum points upwards. The theoretical model dynamics are \cite{Xu2023RobustSeeking}\cite{ApkarianStudentCriteria}:
\begin{equation}
\label{ModelDynamic1}
    M\Ddot{\theta}=m_{p}lr\Ddot{\alpha}\cos\alpha-J_{p}\Dot{\theta}\Dot{\alpha}\sin2\alpha-m_{p}lr\Dot{\alpha}^{2}-b_{r}\Dot{\theta}+\tau
\end{equation}
\begin{equation}
\label{ModelDynamic2}
J_{p}\Ddot{\alpha}=m_{p}lr\Ddot{\theta}cos{\alpha}+0.5J_{p}\Dot{\theta}^{2}\sin2\alpha+m_{p}gl\sin\alpha-b_{p}\Dot{\alpha}
\end{equation}
where \(M=J_{r}+J_{p}\sin^{2}\alpha\) and \(\tau=k_{m}/R_{m}(u-k_{m}\Dot{\theta})\). When the pendulum is balanced and inverted, \(\alpha\) is close to 0. Using small angle approximation. The continuous state space representation of the systems is acquired: 
\newline

$A_{s} =\begin{bmatrix}
0 & 0 & 1 & 0 \\
0 & 0 & 0 & 1 \\
0 & \frac{m_{p}^{2}l^{2}rg}{J_{t}} & -\frac{J_{p}b_{r}}{J_{t}}-\frac{k_{m}^{2}J_{p}}{J_{t}R_{m}} & -\frac{m_{p}lrb_{p}}{J_{t}}\\
0 & \frac{J_{r}m_{p}gl}{J_{t}} & -\frac{m_{p}lrb_{r}}{J_{t}}-\frac{k_{m}^{2}m_{p}lr}{J_{t}R_{m}} & -\frac{J_{r}b{p}}{J_{t}}
\end{bmatrix}$
\newline

$B_{s}=\begin{bmatrix}
0 \\
0\\
\frac{k_{m}J_{p}}{R_{m}J_{t}}\\
\frac{k_{m}m_{p}lr}{R_{m}J_{t}}\\
\end{bmatrix}$
\newline

Where \(J_{t}=J_{p}J_{r}-m_{p}^2l^{2}r^{2}\). All relevant parameters are provided by the Quanser company \cite{ApkarianStudentCriteria}, which are listed in Table \ref{tab:modelparameters1}.
The linear MPC assumes a discrete state space model, where $A=e^{A_{s}T}$, $B=\int_{0}^{T}e^{A_{s}\tau}B_{s}d\tau$, and $T$ refers to time step.

\begin{table}[!t]
\caption{QUBE-Servo 2 Rotary Pendulum Parameters\label{tab:modelparameters1}}
\centering
\begin{tabular}{c||c||c}
\hline
Symbol & Description & Value\\
\hline
\(R_{m}\) & Terminal Resistance & 8.4 \(\Omega\)\\
\(k_{t}\) & Current-torque & 0.042 \(N\cdot m/A\)\\
\(k_{m}\) & Back-emf Constant & 0.042 \(V\cdot s/rad\)\\
\(m_{r}\) & Rotary Arm Mass & 0.095 \(kg\)\\
\(r\) & Rotary Arm Length & 0.085 \(m\)\\
\(b_{r}\) & Rotary Arm Damping Coefficient & 0.001 \(N\cdot m\cdot s/rad\)\\
\(m_{p}\) & Pendulum Mass & 0.024 \(kg\)\\
\(2l\) & Pendulum Length & 0.129 \(m\)\\
\(b_{p}\) & Pendulum Damping Coefficient & 0.00005 \(N\cdot m\cdot s/rad\)\\
\hline
\end{tabular}
\end{table}

\subsection{MPC design and NNMPC training}
Using the MATLAB mpcDesigner Toolbox, a constrained and stable MPC framework was built based on the design criteria in \cite{Mayne2000ConstrainedOptimality}, where the weight is defined as $q_{1}=5,q_{2}=5,q_{3}=0,q_{4}=0,r_{1}=0.5$, the constraints are defined as $|x_{1}|<=2,|r_{1}|<=12$, the prediction horizon $N$ is 50, and the sample time $Ts$ is 0.01s. 

The MPC algorithm not only takes in the current states but also all the reference steps within the prediction horizon for each state, implying that it will require $n + n*N$ neurons for the input layer. In order to reduce the complexity of the NN architecture and the computational cost of the NNMPC, a technique which can help "shrink" the prediction steps is proposed. Instead of taking in all the prediction steps for NNMPC training, we downsample some crucial points which can represent the general path within the prediction horizon. This prediction part for the NNMPC input layer is particularly effective for reference tracking control scenario. Assuming a trajectory $f_{ref}(t) \in\mathbb{R}^n$ that the MPC is currently tracking, the reference part for the MPC at current time $k$ is $W = [f_{ref}(k),f_{ref}(k+T_s),f_{ref}(k+2*T_s),...,f_{ref}(k+(N-1)*T_s)]$, which has a size of $N$. Now we downsample some crucial points $C$ at $f_{ref}(t)$ by increasing the sample time $T_s$ to $Cts$ (note that $CT_s$ needs to be divisible to $T_s$, and $N$ need to be divisible to $CT_s/T_s$), then $C = [f_{ref}(k),f_{ref}(k+CT_s),f_{ref}(k+2*CT_s),...,f_{ref}(k+(N*T_s/CT_s)*CT_s)]$, which has a size of $N/(CT_s/T_s)+1$. 

\begin{figure}[h]
\centering
\includegraphics[width=3.49in]{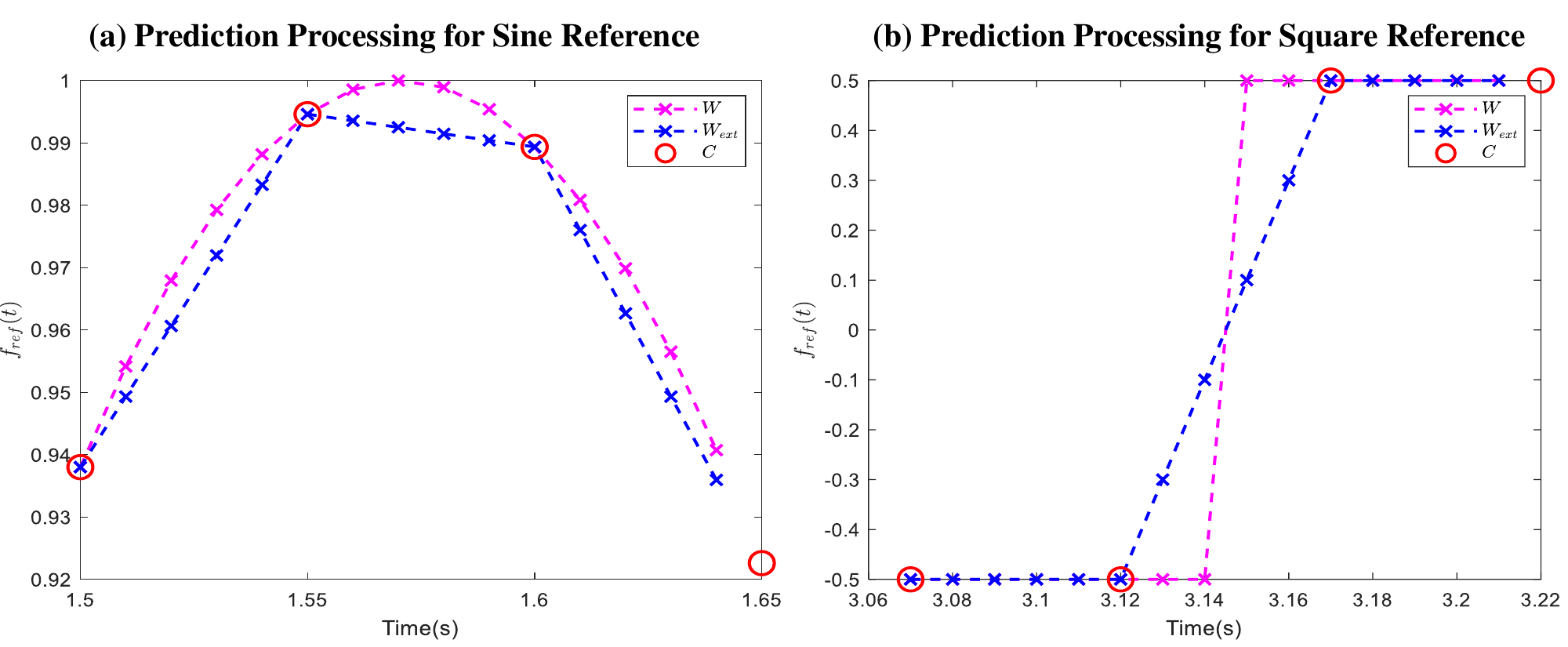}
\caption{An example of prediction steps processing using the reference "shrinking" downsampling technique. (a) is a sine wave with a period of 0.4*pi and an amplitude of 1. (b) is a square wave with a period of 0.4*pi and an amplitude of 0.5}
\label{fig_prediction}
\end{figure}

These downsampled points will serve as the reference for the NNMPC training and will be used to regenerate the all prediction steps for MPC using extrapolation. Thus processed reference for MPC is $W_{ext} = [f_{ref}(k),f_{ref}(k+Ts),f_{ref}(k+2*Ts),...,f_{ref}(k+(N-1)*Ts)]$. In this paper, the MPC using this technique is called Downsampled Reference Model Predictive Control (DRMPC).

Compared to the original one, the size of the input layer is reduced to a great extent when the prediction step is large. However, some minor deficiencies could happen since the reference steps MPC take in are not a prefect representation of the reference trajectory. In this effort, the reliability of the technique was tested in simulation for our system first before using DRMPC in later steps - in future efforts, a closer analysis of the general implications of this reference downsampling is justifiable.

With this technique in hand, the training set can be generated, where the input for the NNMPC is $[\bm{x},C]$, and output will be result generated by MPC using model state $\bm{x}$ and processed reference $W_{ext}$, which is $f_{mpc}(\bm{x},W_{ext})$. In the context of Qube-Servo 2, the general control is to rotate the rotary arm while keeping the inverted pendulum balanced, so the reference of $x_{2}$ and $x_{4}$ is always zero. Since the weight for $x_{3}$ is zero, and it will be not counted as a part of cost function, so the reference for $x_{3}$ would not influence the control performance. Thus, in NNMPC training, we only need the reference trajectory for $x_{1}$, so the input layer only need $[\bm{x},C(1,:)]$, where $C(1,:)=[c_{1},c_{2},c_{N/(Cts/Ts)+1}]$ contains all the sampled crucial point for $x_{1}$, reducing the required neurons again. 

\subsection{RLMPC Design}
In this part, we combine NNMPC and DDPG policy to get a new predictive control algorithm. The actor network, updated through the reward function, is used as an input corrector of the NNMPC, improving the overall control performance. The reward function of the RL environment is defined as:
\begin{equation}
\label{Reward function}
    \text{Reward}=-5(x_{1}-c_{1})^{2}-5(x_{2})^{2}-0.5(u_{1})^{2}
\end{equation}
where the $c_{1}$ is the current reference step for $x_{1}$. The reward function aims to approximate the cost function of the MPC used for training NNMPC. The voltage constraint for QUBE-Servo 2 is defined as $\pm15V$. To achieve the optimization goal while not violating the voltage constraint, we propose two RLMPC variations: 

\subsubsection{Warm Start RL}
This approach will initialize the actor network with the pre-trained NNMPC. Thus the architecture of the actor network should match the NNMPC settings. The action bound for the actor network is defined as $\pm15V$, the voltage applied to the plant is: 
\begin{equation}
\label{WarmStartRLMPC}
    \bm{u}_{k}'=\mu(\bm{s}_{k}|\theta^{\mu})
\end{equation}
where the $\bm{s}_{k} = [\bm{x}_{k},C_{k}(1,:)]$ .$\bm{u}_{k}'$ is optimized to $\bm{u}_{k}^{*}$ which maximize the cumulative reward. The weight $\mu$ at first is initialized as the weight of NNMPC $\pi_{nnmpc}$, and then converge to $\mu^{*}$ during training based on the defined reward function. 

\subsubsection{RL + MPC}
\begin{figure}[h]
\centering
\includegraphics[width=3.49in]{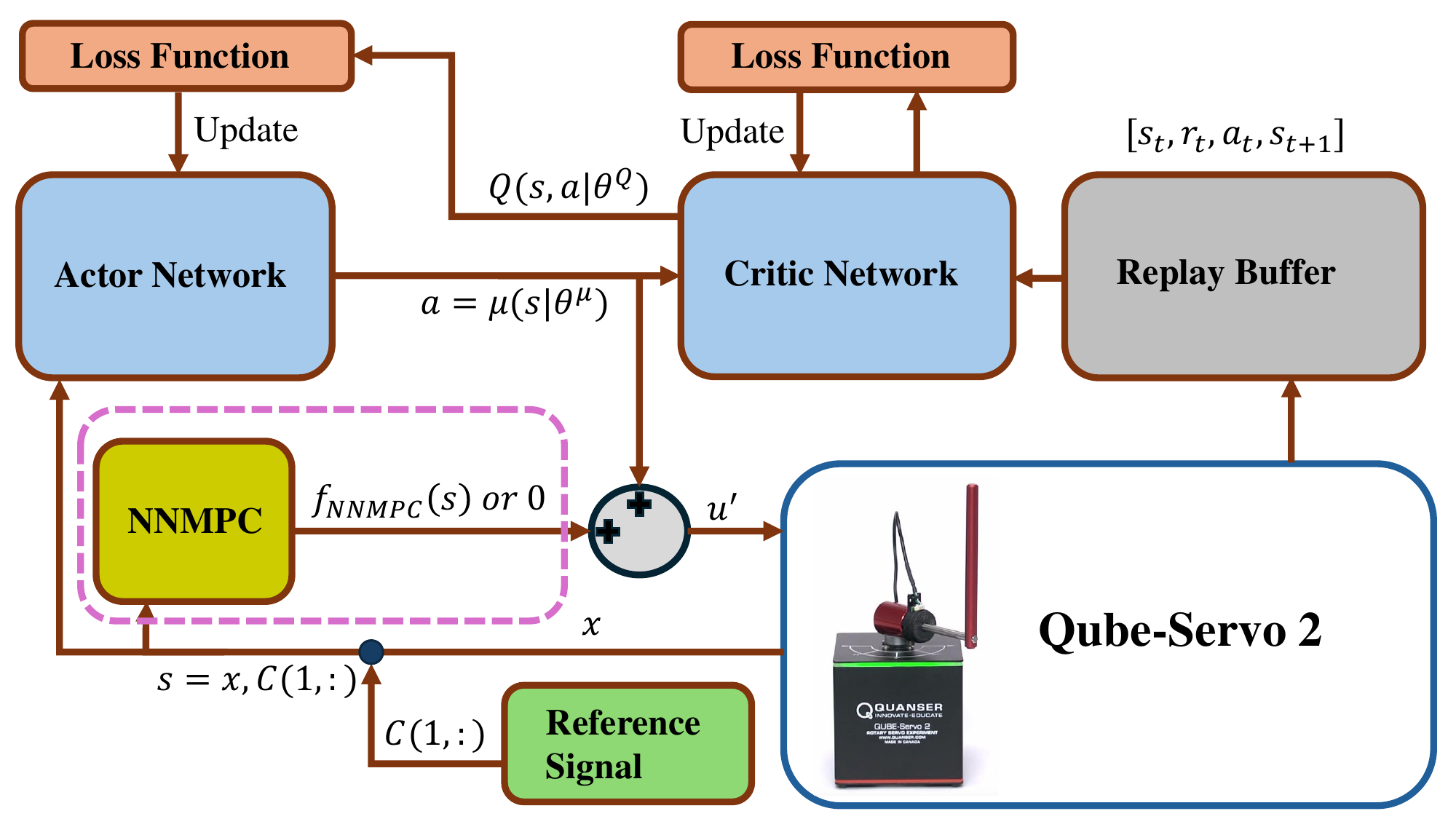}
\caption{A diagram illustrating the logic of RLMPC. When using approach 1): Warm Start RL, the NNMPC part in the picture will be waived and Actor Network will be initialized as an NNMPC. When using approach 2): RL + MPC, the input for the plant is the addition of MPC output and actor network output.}
\label{fig_1}
\end{figure}

This approach will separate NNMPC and actor network. Considering safety as one of the crucial factors, we let the NNMPC take most part of the control and let RL policy do the minor voltage correction during each training episode. The aggressive action led by the exploration characteristics of the actor network can be compensated by the feedback loop of NNMPC within a certain range of parameter variation. The NNMPC input bound is $\pm12V$ and actor network input bound is $\pm3V$. The total voltage input is calculated as $\bm{u}_{k}'$: 
\begin{equation}
\label{RLMPCinput}
    \bm{u}_{k}'=f_{MPC}(\bm{x}_{k},C_{k}(1,:))+\mu(\bm{s}_{k}|\theta^{\mu})
\end{equation}
where $\bm{u}_{k}'$ is optimized to $\bm{u}_{k}^{*}$ as $\mu$ converge to $\mu^{*}$ during training based on the defined reward function. The overall framework is shown in Figure \ref{fig_1}.

From Figure \ref{fig_1} it can be seen that the actor network is updated using the action value generated by critic network. Thus, a reliable critic network is necessary for successful training result. The critic network is updated using the data collected in the replay buffer. But at the begining of the training, the replay buffer is not big enough and the critic network need some time to converge so the critic network is giving unreliable action value. The actor network will be updated using the "immature" action value generated by critic network, leading to suboptimal performance and loss of stability. Thus, it is more practicable to train a critic network in simulation with theoretical dynamic model and initialize the critic network with the pre-trained one in real world training scenarios. 

To get successful training results, training dataset with different characteristics is needed. That is, the replay buffer must have enough observations with different features in transitions set, thus the prediction reference crucial steps which was treated as part of the state observation have to be unique enough so that a effective actor network is trained. 

\begin{algorithm}[H]
\caption{RLMPC Training}\label{RLMPCalgo}
\begin{algorithmic}
\STATE Initialize networks and training parameters
\STATE Initialize maximum training step T and episode number M
\STATE \textbf{for} episode i=1:M 
\STATE \hspace{0.5cm}Randomly initialize reference $f_{ref}(t)$
\STATE \hspace{0.5cm}Initialize reference signal $C_{1}$ and reset state $\bm{x}_{1}$
\STATE \hspace{0.5cm}\textbf{for} time step k=1:T 
\STATE \hspace{0.5cm}\hspace{0.5cm}Generate input based on current policy:
\STATE \hspace{0.5cm}\hspace{0.5cm}\textbf{if} RLMPC == Warm Start RL
\STATE \hspace{0.5cm}\hspace{0.5cm}\hspace{0.5cm}$\bm{u}_{k}'=\mu(\bm{x}_{k},C_{k}|\theta^{\mu})+\mathcal{N}$
\STATE \hspace{0.5cm}\hspace{0.5cm}\textbf{else} 
\STATE \hspace{0.5cm}\hspace{0.5cm}\hspace{0.5cm}$\bm{u}_{k}'=f_{nnmpc}(\bm{x}_{k},C_{k})+\mu(\bm{x}_{k},C_{k}|\theta^{\mu})+\mathcal{N}$
\STATE \hspace{0.5cm}\hspace{0.5cm}Update model state $\bm{x}_{k+1}=\hat{f}_{model}(\bm{x}_{k},\bm{u}_{k}')$
\STATE \hspace{0.5cm}\hspace{0.5cm}Observe reward $r_{k}$ and intercept $C_{k+1}$ from $f_{ref}(t)$
\STATE \hspace{0.5cm}\hspace{0.5cm}Store transition $(\bm{s}_{k},\bm{a}_{k},\bm{r}_{k},\bm{s}_{k+1})$ in replay buffer
\STATE \hspace{0.5cm}\hspace{0.5cm}Update the actor and critic using policy gradient
\STATE \hspace{0.5cm}\hspace{0.5cm}\textbf{if} $\bm{x}_{k+1}$violates the constraint $\bm{x}_{max}$, $\bm{x}_{min}$
\STATE \hspace{0.5cm}\hspace{0.5cm}\hspace{0.5cm}\textbf{break}
\STATE \hspace{0.5cm}\textbf{end for}
\STATE \textbf{end for}  
\end{algorithmic}
\label{alg1}
\end{algorithm}

In real world RLMPC training situations, Algorithm \eqref{alg1} needs to be modified to reset the state of the model and handle the constraint to ensure the training safety. To address this question, we introduce several different constraints with different control methods. During real world training, due to the uncertainty in model and the aggressive policy gradient of actor network, the resulting state $x$ may violate the hard constraints of the actual model or the NNMPC. Thus, we proposed a soft constraint that can be violated during the training. When violating the soft constraint because of RLMPC, the training will be disabled and an offline NNMPC algorithm will be used to reset the state and stabilize the system. Once the system is stabilized and the states are in the reset range, start a new training task.

\begin{figure}[h]
\centering
\includegraphics[width=3.2in]{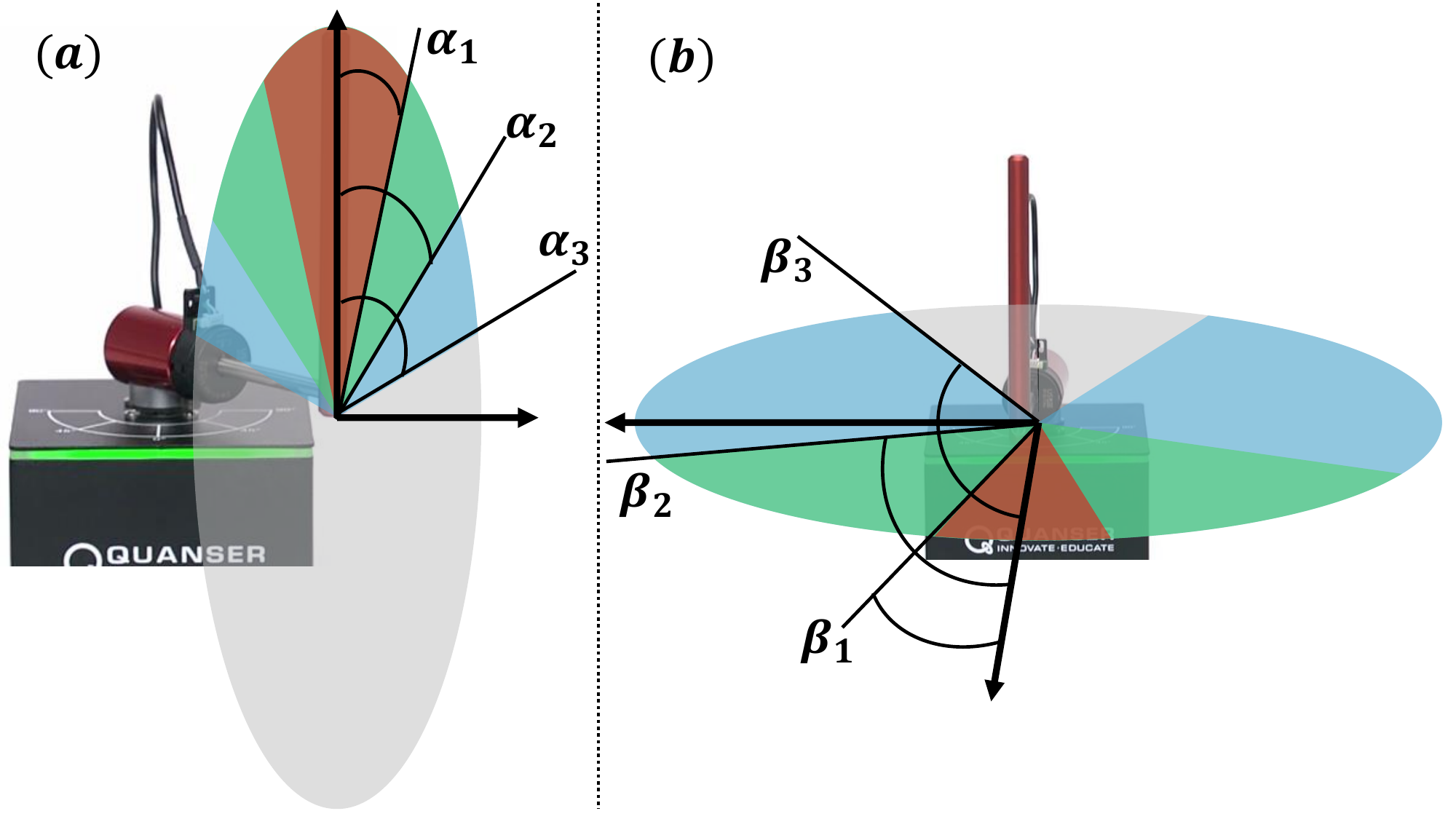}
\caption{This is a diagram for the defined constraints, (a) for pendulum angle $x_{2}$, (b) for rotary angle $x_{1}$. Where $|x_{2}|<|\alpha_{1}|$ and $|x_{1}|<|\beta_{1}|$ are the reset range, $|x_{2}|<|\alpha_{2}|$ and $|x_{1}|<|\beta_{2}|$ are soft constraints, $|x_{2}|<|\alpha_{3}|$ and $|x_{1}|<|\beta_{3}|$ are hard constraints. Other constraints incorporating velocities $x_{3}$ and $x_{4}$ like energy equation could be introduced to make the training safer.}
\label{fig_2}
\end{figure}

Considering the physical constraints of the Qube-Servo 2 and NNMPC's ability to stabilize the system within the hard constraint, the constraints have to be carefully designed. When violating the soft constraint during training, a large penalty (negative reward) will be applied: $Penalty > |-5(\alpha_{2}-c_{1})^{2}-5(\beta_{2})^{2}|$. In this experiment, the constraints and the  are summarized in Table \ref{constraints}.

\begin{table}[!t]
\caption{Defined Constraints and Penalty\label{constraints}}
\centering
\begin{tabular}{c||c|c|c|c|c|c||c}
\hline
Constraint & $\alpha_{1}$ & $\alpha_{2}$ & $\alpha_{3}$ & $\beta_{1}$ & $\beta_{2}$ & $\beta_{3}$ & Penalty (-Reward)\\
\hline
Angle (rad)& 0.1 & 0.4 & 0.5 & 0.2 & 1.5 & 2 & 1000\\
\hline
\end{tabular}
\end{table}

For the evaluation of the optimized performance, a average cost function $J_{ac}$ was designed:
\begin{equation}
\label{AverageCost}
    J_{ac}=\frac{1}{t_{f}-t_{s}}\int_{t_{s}}^{t_{f}}5(x_{1}-w_{1})^{2}+5(x_{2})^{2}+0.5(u_{1})^{2}dt
\end{equation}
The average cost function will measure the average cost within a period, providing a criteria for justifying control performance.

\section{Results}
In this part, the comparison between MPC, DRMPC, NNMPC and the results of RLMPC training in both simulation and real-world control were summarized. For training a NNMPC as well as a RL policy network, a Feedforward Neural Network was used, and the network structure is simple: a input layer contains 10 features (4 states and 6 sample points on the reference trajectory used to reveal the MPC prediction steps), a hidden layer contains 128 neurons with ReLU (Rectified Linear Unit) activation function, 1 neuron with $Tanh$ (Hyperbolic Tangent) activation function, followed by scaling with the output bound which works as a constraint. For real-world control and training, one important factor is the runtime. The RL networks are updated within the time period of each control step, that is, within the control sample time: $Ts$, the CPU or GPU need to have enough computational resource to update the networks. Thus, the computational cost for the training could be too huge for a home PC or some medium-performance computers, the computational ability should be pre-justified before applying the RLMPC training algorithm to the actual training part. 

\subsection{MPC, DRMPC, NNMPC Comparison}
\begin{figure}[h]
\centering
\includegraphics[width=2.8in]{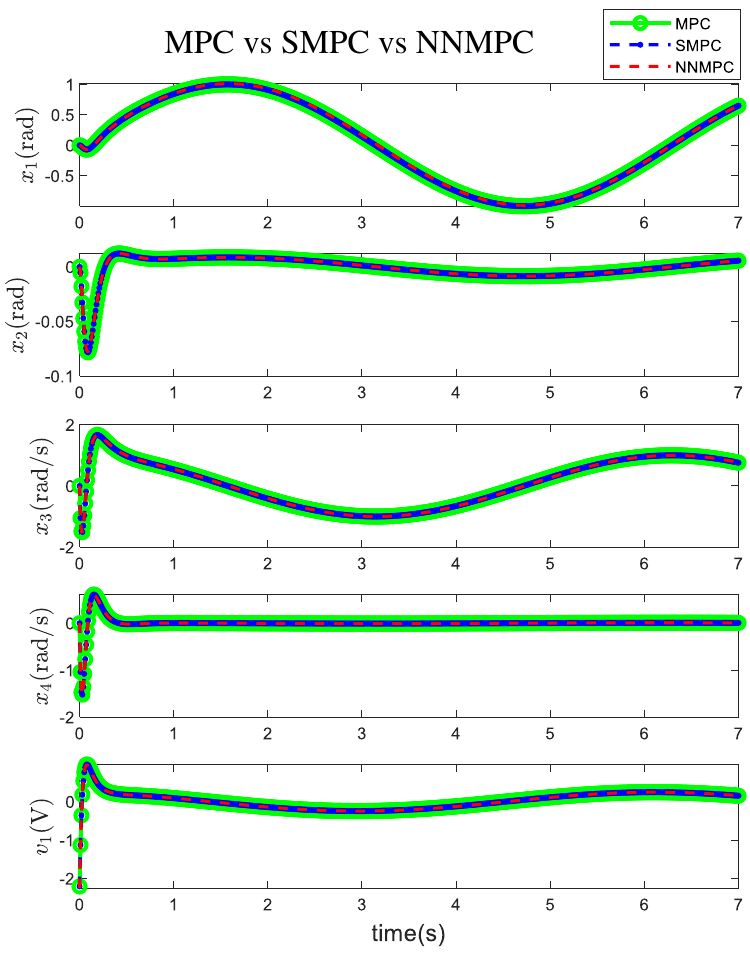}
\caption{An example plot illustrating the differences between MPC, DRMPC, NNMPC in simulation, where the reference is $sin(t)$ and the initial condition is $\bm{x} = [0,0,0,0]$}
\label{fig_simulationMpc}
\end{figure}

This subsection will compare the control performance of MPC, DRMPC, and NNMPC in simulation as well as in real-world control: 

Assuming perfect model with no parameter variations and uncertainties, under different type of reference trajectories, the average cost function (\ref{AverageCost}) was used in the evaluation of the performance of different MPC-based control algorithms: MPC, DRMPC, NNMPC. The results are shown in Table \ref{NNMPCSimulation}. Figure \ref{fig_simulationMpc} gives an example plot of the simulation result.

\begin{table}[ht]
\caption{Average Cost Evaluation for MPC, DRMPC, and NNMPC}
\label{NNMPCSimulation}
\centering
\begin{tabular}{|c|c|c|c|c|c|}
\hline
     Model & MPC &\multicolumn{2}{c|}{DRMPC}&\multicolumn{2}{c|}{NNMPC}\\
\hline
    $f_{ref}$ & $J_{ac}$ & $J_{ac}$ & $\%\uparrow$ & $J_{ac}$ & $\%\uparrow$ \\
\hline
$sin(t)$ & 3.0908 & 3.0875 & -0.11 & 3.1414 & 1.64\\

$0.8*sin(t)$ & 1.9781 & 1.9760 & -0.11 & 2.0576 & 4.02\\

$sin(2t)$ & 12.1101 & 12.0589 & -0.42 & 11.8639 & -2.03\\
\hline
\end{tabular}
\end{table}

Given the comparison of different MPCs under different reference trajectory, it can be seen that the difference between DRMPC, NNMPC and MPC is very small, implying the reliability of the prediction horizon downsampling technique used in DRMPC under different possible reference trajectory type. In addition, the NNMPC approximate the performance of MPC to a great extent while reducing the neuron required for the network structure.

\subsection{RLMPC}
This part compares the performance of original MPC, NNMPC, and trained RLMPCs (both Warm Start RL and RL + MPC) in real-world scenario. To get successful trained networks, the training parameters has to be carefully designed, reviewed and adjusted based on the training results. Based on experiences, the learning rate for actor/critic networks, buffer size are the main factors for getting a satisfying training result. During training, several undesired situations might be observed: RL policy diverges, RL policy converge to a local critic point and get stuck in a suboptimal performance. Considering the durability of the hardware setting, the experiment is firstly done in simulation (not only to get a reliable critic network, but also to confirm if the parameter setting is good enough for the RL policy network to converge), and then done in real world.

\begin{figure}[h]
\centering
\includegraphics[width=3in]{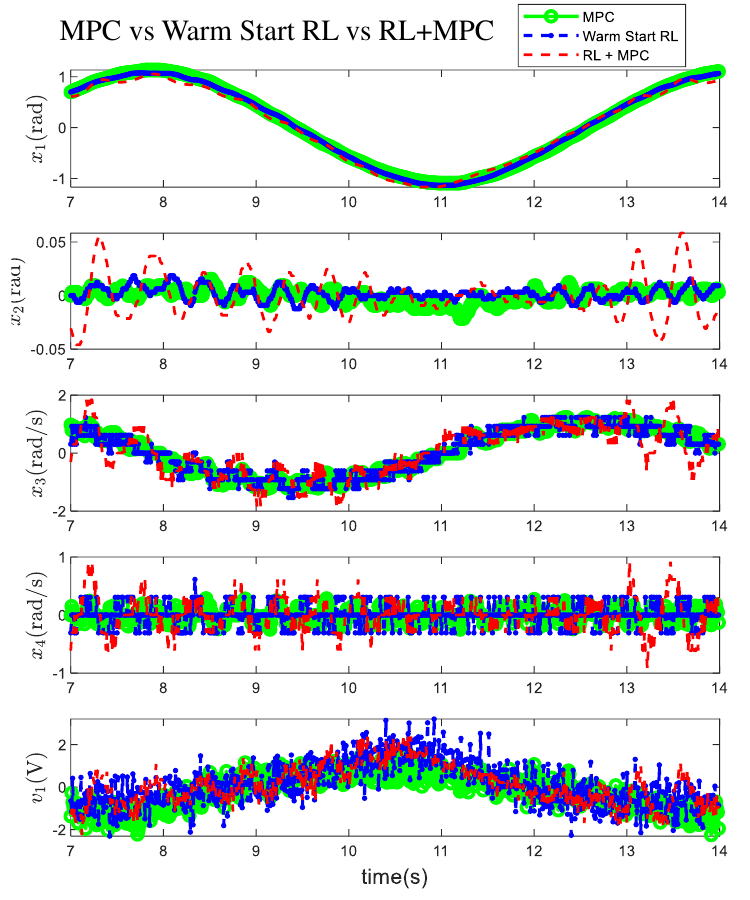}
\caption{An example plot illustrating the control performance of MPC, Warm Start RL, RL+MPC in real world control scenario, where the reference trajectory is $\sin(t)$ and the time starts from $7s$ to $14s$ when the pendulum is stabilized}
\label{fig_RWRLMpc}
\end{figure}

Having designed the training parameters based on the results in simulation, the real world training is performed and the data was gathered. One example plot is shown in Figure \ref{fig_RWRLMpc}.

Based on the data collected in Figure \ref{fig_RWRLMpc}, the average cost for different control algorithms within the $7s-14s$ period is calculated and listed in Table \ref{RLMPCRealWorldComparison}. Five trials were conduct for each $f_{ref}$, Control Algorithm combination, and their average value was recorded.

\begin{table}[ht]
\caption{Average Cost Evaluation for MPC and RLMPCs \label{RLMPCRealWorldComparison}}
\centering
\begin{tabular}{|c|c|c|c|c|c|c|c|}
\hline
     Model & MPC& \multicolumn{2}{c|}{NNMPC} &\multicolumn{2}{c|}{Warm Start RL}&\multicolumn{2}{c|}{RL + MPC}\\
\hline
    $f_{ref}$ & $J_{ac} $& $J_{ac} $ &$\%\uparrow$ & $J_{ac}$ & $\%\downarrow$ & $J_{ac}$ & $\%\downarrow$ \\
\hline
$0.5\sin(t)$ & 27.05& 28.79 & 6.39 & 26.78 & 1.00 & 16.24 & 39.97\\

$0.8\sin(t)$ & 32.91& 34.70 & 5.43 & 31.97 & 2.86 & 26.20 & 20.40\\

$\sin(t)$    & 38.39& 40.27 & 4.91 & 36.53 & 4.83 & 33.34 & 13.14\\
\hline
$1.2\sin(t)$ & 46.07& 47.01 & 2.05 & 44.30 & 3.84 & 40.41 & 12.28\\

$1.4\sin(t)$ & 57.93& 59.91 & 3.43 & 55.04 & 4.98 & 51.24 & 11.54\\
\hline
\end{tabular}
\end{table}

From Table \ref{RLMPCRealWorldComparison} it can be seen that both the Warm Start RL and RL + MPC approaches can reduce the cost, thus mitigating the MPC suboptimal performance problem to a certain extent. One remarkable discovery is that, during training, only $\sin(t)$ reference signals with an amplitude of 1 were used, but RLMPC still works well and making improvement on the overall performance for reference signals such as $1-1.5\sin(t)$ (scaling and offset), implying the neuron network's generalizability and RLMPC's potential practicality for solving further problems.
 
Compared to the NNMPC that approximates the MPC control algorithm, the trained RLMPC appears to create a response that fluctuates around the reference trajectory, creating a smaller cost. The NNMPC is more likely to create an offset between the actual path and trajectory in real-world control scenario, and that offset leads to a bigger cost. However, with different training setting, different number of training episodes, the result may vary.

\begin{table}[ht]
\caption{RLMPC Runtime Analysis \label{MPCRuntime}}
\centering
\begin{tabular}{|c|c|c|c|c|c|c|}
\hline
      MPC& \multicolumn{2}{c|}{NNMPC} &\multicolumn{2}{c|}{Warm Start RL}&\multicolumn{2}{c|}{RL + MPC}\\
\hline
    $t_{1}(\mu s)$& $t_{2}(\mu s)$ &$\%\downarrow$ & $t_{3}(\mu s)$ & $\%\downarrow$ & $t_{4}(\mu s)$ & $\%\downarrow$ \\
\hline
220.2& 0.0838 & 99.962 &0.0850 & 99.961 & 0.2513 & 99.896\\
\hline
\end{tabular}
\end{table}

According to the results in Table \ref{MPCRuntime}, both NNMPC and RLMPCs significantly reduces the computational cost of the original MPC that requires quadratic programming.

\begin{figure}[h]
\centering
\includegraphics[width=3.49in]{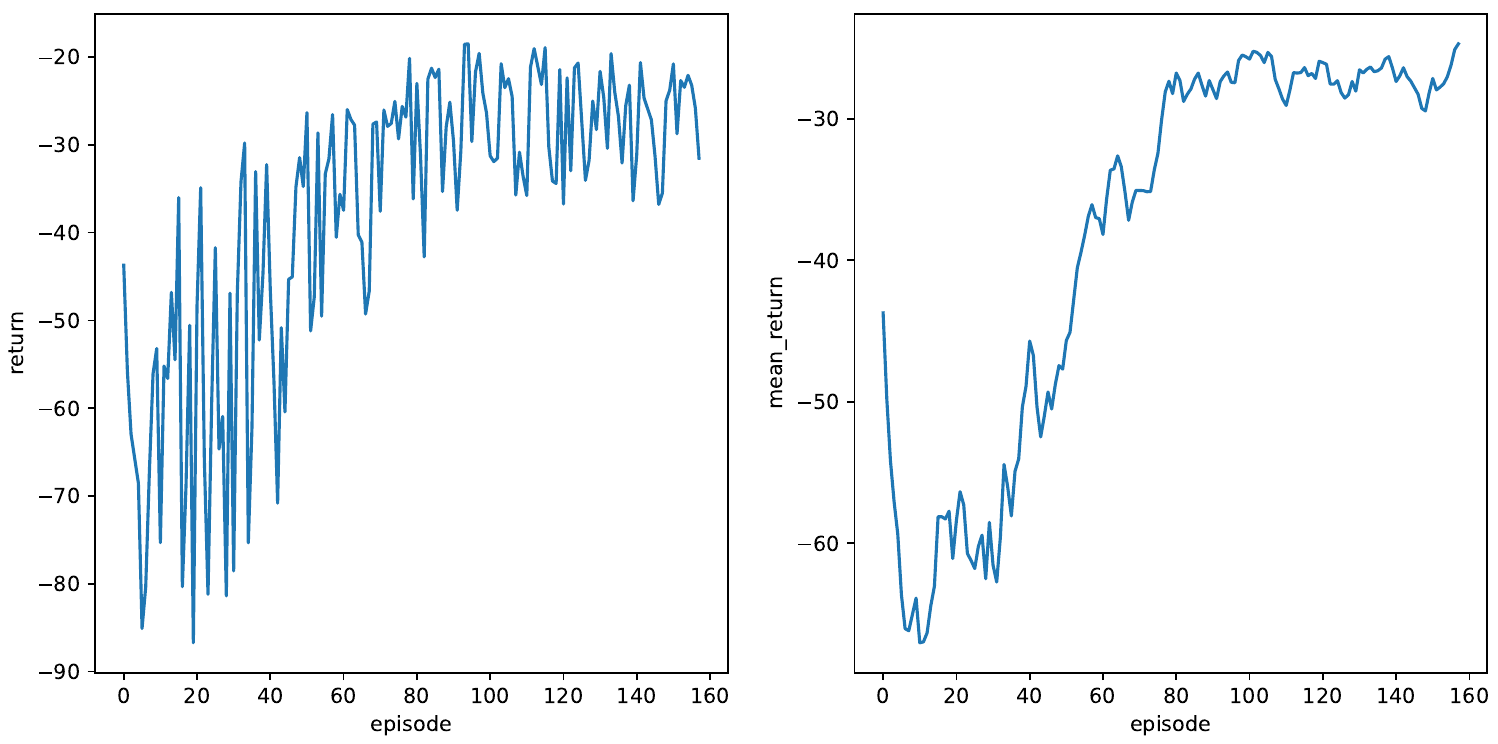}
\caption{An example illustrating the general trend that return and mean improve with increasing training episodes in a successful real-world training}
\label{fig_TrainRLMpc}
\end{figure}

During training, a general trend of episode reward is seen in Figure \ref{fig_TrainRLMpc}. At beginning, the training diverges, which is likely due to the limited data gathered in the buffer. Having enough different datasets in the buffer, it will converge to an improved control algorithm that maximize the reward for each episode. With appropriate training setting, the training may converge within 300 episodes (each episode last for 7s), and so this method could be useful in practice in some types of systems.

Given the data and plot above, the RLMPC approach, with the RL contributions safely constrained, can mitigate MPC challenges with modeling uncertainty while decreasing the computational cost compared to the original MPC algorithm. In the implementation here for the inverted pendulum system, the adaptation occurs over \~100 or so episodes before convergence. The safety issue is addressed by the soft/hard constraint design shown in Figure \ref{fig_2} as well as limiting the maximum possible contribution of the RL portion. This corresponds to the RL additions fitting within some robustness bounds of the original system. While in real world training the model did not violate the physical constraints of the model for this system, future efforts in this direction should consider, more analytically if possible, convergence, stability, and robustness guarantees for the different approaches. The neural network approach in particular can make proving guarantees a particular challenge.

\section{Conclusion}
In this work, we proposed a novel real world Reinforcement Learning based Model Predictive Control framework to solve the control performance challenges by model parameter variations and uncertainties arising from model/actual system mismatch. This approach leverages a neural network to serve as an explicit MPC-like surrogate, to substantially reduce algorithm runtimes. We apply a downsampling method on the reference inputs over the prediction horizon, and consider two approaches to RLMPC.

These two main RLMPC frameworks, one with Warm Start RL tp initialize the actor network with the NNMPC pre-trained as an explicit MPC-like neural network. In a separate approach, the RL+NNMPC approach separates NNMPC and the actor network. In this case, the offline NNMPC will do most part of the control and let RL do the minor voltage correction during each training episode. The aggressive action led by the exploration characteristics of the actor network can be compensated by the feedback loop of NNMPC, perhaps assuring more safety, though this assumption needs to be studied further in future work.

This work points to an interesting direction of future study of stability, performance, and robustness assurance in potential architectures for combining MPC and Reinforcement Learning that can leverage the benefits of both approaches.

\bibliographystyle{IEEEtran}
\bibliography{references}

\begin{thebibliography}{10}
\providecommand{\url}[1]{#1}
\csname url@samestyle\endcsname
\providecommand{\newblock}{\relax}
\providecommand{\bibinfo}[2]{#2}
\providecommand{\BIBentrySTDinterwordspacing}{\spaceskip=0pt\relax}
\providecommand{\BIBentryALTinterwordstretchfactor}{4}
\providecommand{\BIBentryALTinterwordspacing}{\spaceskip=\fontdimen2\font plus
\BIBentryALTinterwordstretchfactor\fontdimen3\font minus \fontdimen4\font\relax}
\providecommand{\BIBforeignlanguage}[2]{{%
\expandafter\ifx\csname l@#1\endcsname\relax
\typeout{** WARNING: IEEEtran.bst: No hyphenation pattern has been}%
\typeout{** loaded for the language `#1'. Using the pattern for}%
\typeout{** the default language instead.}%
\else
\language=\csname l@#1\endcsname
\fi
#2}}
\providecommand{\BIBdecl}{\relax}
\BIBdecl

\bibitem{Lin2023ReinforcementSystems}
M.~Lin, Z.~Sun, Y.~Xia, and J.~Zhang, ``{Reinforcement Learning-Based Model Predictive Control for Discrete-Time Systems},'' \emph{IEEE Transactions on Neural Networks and Learning Systems}, 2023.

\bibitem{Martin2017SensitivityDrives}
\BIBentryALTinterwordspacing
C.~Martin, M.~B. Guzman, F.~Barrero, M.~R. Arahal, X.~Kestelyn, M.~J. Duran, C.~Mart{\'{i}}n, M.~Berm{\'{u}}dez, and M.~J. Dur{\'{a}}n, ``{Sensitivity of predictive controllers to parameter variation in five-phase induction motor drives},'' Tech. Rep., 2017. [Online]. Available: \url{https://hal.science/hal-01899200}
\BIBentrySTDinterwordspacing

\bibitem{Bujarbaruah2018AdaptiveKeeping}
\BIBentryALTinterwordspacing
M.~Bujarbaruah, X.~Zhang, H.~E. Tseng, and F.~Borrelli, ``{Adaptive MPC for Autonomous Lane Keeping},'' 6 2018. [Online]. Available: \url{http://arxiv.org/abs/1806.04335}
\BIBentrySTDinterwordspacing

\bibitem{Borrelli2017PredictiveSystems}
F.~Borrelli, A.~Bemporad, and M.~Morari, \emph{{Predictive Control for Linear and Hybrid Systems}}, 2017.

\bibitem{Gros2022LearningGuarantees}
S.~Gros and M.~Zanon, ``{Learning for MPC with stability {\&} safety guarantees},'' \emph{Automatica}, vol. 146, 2022.

\bibitem{Brunton2019Data-DrivenControl}
S.~L. Brunton and J.~N. Kutz, \emph{{Data-Driven Science and Engineering: Machine Learning, Dynamical Systems, and Control}}, 2019.

\bibitem{Hardle2019AppliedAnalysis}
W.~K. H{\"{a}}rdle and L.~Simar, \emph{{Applied Multivariate Statistical Analysis}}, 2019.

\bibitem{Wong2018RecurrentManufacturing}
\BIBentryALTinterwordspacing
W.~C. Wong, J.~Li, and X.~Wang, ``{Recurrent Neural Network-based Model Predictive Control for Continuous Pharmaceutical Manufacturing},'' 7 2018. [Online]. Available: \url{http://arxiv.org/abs/1807.09556}
\BIBentrySTDinterwordspacing

\bibitem{Karg2020EfficientLearning}
B.~Karg and S.~Lucia, ``{Efficient Representation and Approximation of Model Predictive Control Laws via Deep Learning},'' \emph{IEEE Transactions on Cybernetics}, vol.~50, no.~9, 2020.

\bibitem{Wang2009ModelMATLAB}
L.~Wang, \emph{{Model Predictive Control System Design and Implementation Using MATLAB{\textregistered}}}, 2009.

\bibitem{Lillicrap2015ContinuousLearning}
\BIBentryALTinterwordspacing
T.~P. Lillicrap, J.~J. Hunt, A.~Pritzel, N.~Heess, T.~Erez, Y.~Tassa, D.~Silver, and D.~Wierstra, ``{Continuous control with deep reinforcement learning},'' 9 2015. [Online]. Available: \url{http://arxiv.org/abs/1509.02971}
\BIBentrySTDinterwordspacing

\bibitem{Xu2023RobustSeeking}
J.~Xu, Y.~Tan, and X.~Chen, ``{Robust Tracking Control for Nonlinear Systems: Performance optimization via extremum seeking},'' in \emph{Proceedings of the American Control Conference}, vol. 2023-May, 2023.

\bibitem{ApkarianStudentCriteria}
J.~Apkarian and Q.~Michel~L{\'{e}}vis, ``{Student Workbook Qube-Servo experiment for MAtLAb /Simulink users Standardized for ABET * Evaluation Criteria},'' Tech. Rep.

\bibitem{Mayne2000ConstrainedOptimality}
D.~Q. Mayne, J.~B. Rawlings, C.~V. Rao, and P.~O. Scokaert, ``{Constrained model predictive control: Stability and optimality},'' \emph{Automatica}, vol.~36, no.~6, pp. 789--814, 6 2000.

\end{thebibliography}

\end{document}